\documentclass[twoside,twocolumn,10pt]{article}
\usepackage{amsmath}
\usepackage{authblk}
\usepackage[T1]{fontenc}
\usepackage[utf8]{inputenc}
\usepackage{textcomp}
\usepackage{amstext}
\usepackage{amssymb}
\usepackage{bm}
\usepackage{amsfonts}
\usepackage{amssymb}
\usepackage{graphicx}
\usepackage[usenames,dvipsnames]{xcolor}
\usepackage{hyperref}

\newcommand{\MoS}[1]{\(\mathrm{MoS_{#1}}\)}
\newcommand{\SiO}[1]{\(\mathrm{SiO_{#1}}\)}
\newcommand{\Si}{\(\mathrm{Si}\)}
\newcommand{\tens}[1]{{\overline{#1}}}
\begin{document}

\title{Electrodynamic models of 2D materials: can we match thin film and single sheet approaches?}

\author[1]{Bruno Majérus}
\author[2]{Evdokia Dremetsika}
\author[1,3]{Michaël Lobet}
\author[1]{Luc Henrard}
\author[2]{Pascal Kockaert\thanks{Corresponding author: \texttt{Pascal.Kockaert@ulb.ac.be}}}
\affil[1]{Department of Physics \& Namur Institute of Structured Matters (NISM), University of Namur, 61 rue de Bruxelles, B-5000 Namur, Belgium.}
\affil[2]{OPERA-photonics, Université libre de Bruxelles (U.L.B.), 50 Avenue F. D. Roosevelt, CP 194/5, B-1050 Bruxelles, Belgium}
\affil[3]{John A. Paulson School of Engineering and Applied Sciences, Harvard University, 9 Oxford Street, Cambridge, MA 02138, United States of America}
\maketitle

\begin{abstract}
The electromagnetic properties of 2D materials are modeled either as single sheets with a surface susceptibility or conductivity, or as thin films of finite thickness with an effective permittivity. Their intrinsic anisotropy, however, has to be fully described to reliably predict the optical response of systems based on 2D materials or to unambiguously interpret experimental data. In the present work, we compare the two approaches within the transfer matrix formalism and provide analytical relations between them. We strongly emphasize the consequences of the anisotropy. In particular, we demonstrate the crucial role of the choice of the thin film’s effective thickness compared with the parameters of the single sheet approach and therefore the computed properties of the 2D material under study.
Indeed, if the \textit{isotropic} thin film model with very low thickness is similar to an \textit{anisotropic} single sheet with no out-of-plane response, with larger thickness it matches with a single sheet with \textit{isotropic} susceptibility, in the reasonable small phase condition.
We illustrate our conclusions on extensively studied experimental quantities such as transmittance, ellipsometry and optical contrast, and  we discuss similarities and discrepancies reported in the literature when using single sheet or thin film models.
\end{abstract}

\section{Introduction}

The electromagnetic (EM) properties of 2D materials are at the forefront of the present research activities. Further developments for applications as diverse as optical modulators, transparent conductive films, photovoltaic systems, superabsorbers or sensors request an accurate description of the electromagnetic response  ~\cite{doi:10.1002/adma.201606128,doi:10.1038/nphoton.2010.186,doi:10.1039/c4nr01600a,doi:10.1038/nphoton.2015.282}. For example, optical contrast or transmission are among the commonly used quantities to characterize 2D systems, in particular, to determine their thickness or their number of layers~\cite{Bayle_18,doi:10.1063/1.2768625,Eichfeld_2014,Li_2013,ottaviano_mechanical_2017}. Furthermore, electromagnetic properties are the macroscopic fingerprints of elementary excitations such as the inter-band transition, excitons or plasmons. Their correct analysis is therefore crucial for the understanding of the underlying physics of 2D materials.

Several models have been recently used in this context. The EM response to an external field has been firstly considered as a purely 2D phenomenon with the definition of a single sheet (surface) conductivity \(\sigma^s\), or susceptibility \(\chi^s\)~\cite{PhysRevB.78.085432,Lobet_16}. In particular, for graphene, an analytical expression for \(\sigma^s\) based on tight-binding approximation and Kubo formula has become popular~\cite{doi:10.1088/1742-6596/129/1/012004} and provides a clear distinction between inter-band and intra-band electronic transitions. The surface conductivity can be determined experimentally, \textit{e.g.} via Brewster angle measurements~\cite{Majerus_18}.

Using another approach, 2D materials have been considered as isotropic materials with a small but finite thickness ~\cite{Eichfeld_2014,Li_2013,ottaviano_mechanical_2017,Nelson_2010,cheon_how_2014}. This approach notably allows to use the well-developed transfer matrix technique  to predict and  interpret optical data (including ellipsometry) with widely available methodology and numerical codes. The thickness is often arbitrarily taken as the interlayer distance of the 3D counterpart of the 2D material~\cite{Lobet_16,Lu_97}, considered as a fitting parameter~\cite{Eichfeld_2014} or evaluated based on the variation of the electronic density in the transverse direction~\cite{Wagner_13}.

However, these two  approaches (a purely 2D surface conductivity and a 3D isotropic thin film)  do not match as demonstrated  analytically and numerically~\cite{PhysRevA.93.013832,Matthes_16} and give model-dependent interpretation of ellipsometric data~\cite{Jayaswal_18}. This is particularly true for oblique incidence and TM (\(p\)-polarised) EM radiation~\cite{Valuev2016}. Indeed, considering only a purely in-plane 2D conductivity means that the out-of-plane response of the layer is neglected, while for an isotropic thin film, both the in-plane and out-of-plane responses are linked. Very recently, a criterion has been proposed to determine in which conditions the two models give similar results at normal incidence~\cite{Li_Heinz_18}.

Anisotropic thin films have also been studied. The out-of-plane component has been taken as a free parameter~\cite{Jayaswal_18,Funke_16,Kravets_10},
or deduced from first principle approach calculations performed with periodic boundary conditions~\cite{Matthes_16, Marinopoulos_04}. The out-of-plane susceptibility in a single sheet model has been recently considered by two of us to analyze the 
non-linear optical response of graphene~\cite{PhysRevB.96.235422}. An adequate description of the out-of-plane component is of prime necessity since very diverse 2D materials with potentially large out-of-plane polarisabilities are synthesized~\cite{Novoselov_16} or predicted~\cite{Rasmussen_15}.

In this work, we study analytically and numerically the conditions on the EM response function (surface susceptibility, dielectric tensor) and on the thickness of the effective thin film for a correct description of the response of 2D materials. In particular, we analytically link the surface susceptibility of the single sheet to the ordinary and extraordinary optical constants of the equivalent thin film. We then focus our attention on the determination of the surface conductivity of the 2D materials based on the interpretation of optical transmission, ellipsometry and optical contrast measurements.

\section{Modelisation of 2D materials}

\begin{figure}\centering
\includegraphics{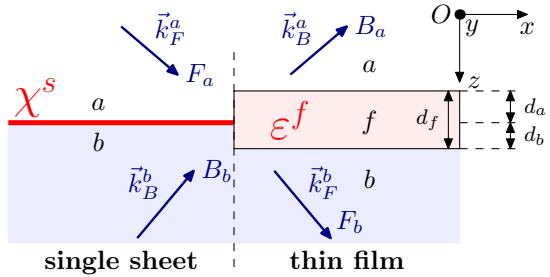}
\caption{Schematic representation of the two configurations. Left: current sheet model. Right: thin film with an effective material \(f\) extending over a distance \(d_a\) (resp. \(d_b\)) on the \(a\) (resp. \(b\)) side.
The wave vectors of the forward \(F\) and backward~\(B\) fields in media \(a\) and~\(b\) are denoted by
\(\vec{k}^{a,b}_{F,B}\).
The reference frame is \(Oxyz\).
}
\label{Fig:interface}
\end{figure}
In this section, we perform the comparison between the single sheet and the thin film approaches within the framework of transfer matrix formalism for stratified media~\cite[Sec.~4.6]{azzam1987ellipsometry}. As a first step, we build the transfer matrix of a single sheet at the interface of two surrounding media (respectively \(a\) and \(b\)), as depicted on Fig.~\ref{Fig:interface} (left). In a second step, we calculate the transfer matrix of a thin film with finite thickness \(d_f\) Fig.~\ref{Fig:interface}  (right). We then analytically compare the two approaches and highlight the consequences on quantities that can be easily determined experimentally (transmittance, ellipsometric data, optical contrast). Importantly, we insist here on the consequences of the intrinsic anisotropy of 2D materials.

The single sheet is described by a surface susceptibility tensor \(\tens{\chi}^s\) diagonal in our reference frame (Fig.~\ref{Fig:interface}). The in-plane components of the 2D material are directly related to its surface conductivity by
\begin{equation}
\sigma^s_{\alpha}=-\mathrm{i}\varepsilon_0\omega\chi^s_{\alpha}
,\label{Eq:sigmas}
\end{equation}
 where \(\alpha=x, y\). The out-of-plane component of the susceptibility \(\chi^s_{z}\) is also considered here, but an out-of-plane conductivity has no physical meaning for a single sheet.

The thin film material is described by a dielectric tensor \(\tens{\varepsilon}^f\) related to the bulk conductivity components by
\begin{equation}
\varepsilon_{\alpha\beta }^f = \left(\varepsilon_0  + \frac{i \sigma_{\alpha}}{\omega}\right)\delta_{\alpha\beta}
.\label{Eq:eps_alphaalpha}
\end{equation}

The in-plane and bulk conductivities are related by \(\sigma^s_{\alpha} = d_f \, \sigma_{\alpha}\)~\cite{Matthes_16}.
The incident and substrate materials (\(a\) and~\(b\)) can be anisotropic but with their optical axes aligned with those of the 2D material, \textit{i.e.} \(\varepsilon_{\alpha \beta}^{a,b}=\varepsilon_0\varepsilon_{\alpha}^{a,b}\delta_{\alpha \beta}\), which is the case in most (if not all) the systems studied experimentally so far. This hypothesis avoids the coupling between transverse electric (TE) and transverse magnetic (TM) modes. We allow those surrounding materials to have a complex permittivity, and express the dependence in the angle of incidence via the wavevector \(\vec{k}= (k_x,k_y,k_z)\). If the incident medium is a perfect dielectric characterized by the real isotropic permittivity \(\varepsilon_0\varepsilon_a=\varepsilon_0n_a^2\), and \(k_y=0\), we have \(k_x=k_0\,n_a\sin\alpha_i\), with \(k_0\) the wavenumber of the light in vacuum,  \(n_a\) the refractive index of medium \(a\), and \(\alpha_i\) the angle of incidence.

\subsection{2D material as a single sheet}
\label{Sec:singlesheet}

In order to describe the out-of plane component in the single sheet model, we use the approach described in~\cite{PhysRevB.96.235422}, based on~\cite{sipe_new_1987,felderhof_electromagnetic_1987}. In particular, the transmission coefficient \(t\) and the reflexion coefficient \(r\) of the electric field in TE and TM configurations can be written as
\begin{eqnarray}
 t &=& t_{ab} \left[1+\mathrm{i}(\varphi_{x}+\varphi_{y}+\varphi_{z})\right],\label{Eq:t}\\
 r &=& t-1 - 2\mathrm{i}\varphi_{x},\label{Eq:r}
\end{eqnarray}
with the parameters defined in table~\ref{Tab:TETM}. We note that \(t_{ab}\), which is the transmission coefficient in absence of 2D material, depends on the propagation direction and is therefore not symmetrical, \textit{i.e.} \(t_{ba}=2-t_{ab}\), while \(\varphi_x,\varphi_y,\varphi_z\) do not depend on the propagation direction.

\begin{table}\centering
\caption{Definition of the coefficients in TE and TM configurations, to the first order in \(\varphi_x,\varphi_y,\varphi_z\). In these expressions, \(m\) and \(n\)  will be replaced by \(a\), \(b\) and \(f\) to denote respectively the incidence medium, the substrate and the thin film. The forward (resp. backward) component \(F_m\) (resp. \(B_m\)) is defined in each medium \(m\) with respect to the forward (resp. backward) component of the electric field parallel to the interface \([E^m_{x,y}]_F\) (resp. \([E^m_{x,y}]_B\) ), with \(y\) for TE, and \(x\) for TM.}
\label{Tab:TETM}
\begin{tabular}{l|cc}
 {} & TE (s-polarization) & TM (p-polarization)\\
 \hline
 \(k_0\)         & \multicolumn{2}{c}{\(\omega/c\)}\\
 \(k_x\)         & \multicolumn{2}{c}{in-plane component of input \(\vec{k}\)}\\
 \(k^m_z\)       & \(\sqrt{\varepsilon^m_y k_0^2-k_x^2}\) & \(\sqrt{\varepsilon^m_x (k_0^2-k_x^2/\varepsilon^m_z)}\)\\
 \(F_m\)           & \([E^m_y]_F\)            & \(\phantom{-}\varepsilon^m_x/k^m_z[E ^m_x]_F\)\\
 \(B_m\)           & \([E^m_y]_B\)            & \(-\varepsilon^m_x/k^m_z[E^m_x]_B\)\\
 \(t\)           & \multicolumn{2}{c}{\(F_b/F_a\)}\\
 \(r\)           & \multicolumn{2}{c}{\(B_a/F_a\)}\\
 \(\alpha_{mn}\) & \(k^n_z /k^m_z\) & \((\varepsilon^m_x k^n_z) / (\varepsilon^n_x k^m_z) \)\\
 \(t_{mn}\)      & \multicolumn{2}{c}{\(2/\left(1+\alpha_{mn}\right)\)}\\
 \(r_{mn}\)      & \multicolumn{2}{c}{\(t_{mn}-1\)}\\
  \(\varphi_x\)   & \(0\)            & \(\frac{k^a_zk^b_z}{\varepsilon^b_x k^a_z + \varepsilon^a_x k^b_z}\chi^s_{x}\)\\ 
 \(\varphi_y\)   & \(\frac{k_0^2}{k^a_z+k^b_z}\chi^s_{y}\)& \(0\)\\ 
 \(\varphi_z\)   & \(0\)            & \(\frac{k^2_x}{\varepsilon^b_x k^a_z + \varepsilon^a_x k^b_z} \frac{\varepsilon^a_x\varepsilon^b_x}{\epsilon_{ab}}\chi^s_{z}\)\\ 
 \(\epsilon_{ab}\) & \multicolumn{2}{c}{\(\frac{2}{\left(1/\varepsilon^a_z+1/\varepsilon^b_z\right)}\)}\\
 \(\varphi_{\pm}\)   & \multicolumn{2}{c}{\(\varphi_x\pm(\varphi_y+\varphi_z)\)}\\
 \(\chi^s_{jj}\) & \multicolumn{2}{c}{\(\mathrm{i}\sigma^s_{jj}/(\varepsilon_0\omega)\)}\\
 \end{tabular}
\end{table}

In these notations the transfer matrix between the incident medium \(a\) and the outgoing medium (substrate) \(b\) can be written as
\begin{equation}
 \mathcal{S}_{ab}
 = \frac{1}{t_{ab}}
 \cdot
 \left( \begin{array}{cc}
 1-\mathrm{i}\varphi_{+}
 &
  r_{ab}+\mathrm{i}\varphi_-
 \\
  r_{ab}-\mathrm{i}\varphi_-
 &
  1+\mathrm{i}\varphi_+
 \end{array}\right),
\label{Eq:matrix2D}
\end{equation}
so that the forward (\(F\)) and backward (\(B\)) field components in media \(a\) and \(b\) at the single sheet interface are linked by
\begin{equation}
 \left(\begin{array}{c} F_a\\ B_a\end{array}\right)
= \mathcal{S}_{ab}
 \left(\begin{array}{c} F_b\\ B_b\end{array}\right).
\label{Eq:FB_S}
\end{equation}
The expressions for TE and TM modes have a similar form if the forward and backward components are defined as in table~\ref{Tab:TETM}.
The matrix \(\mathcal{S}_{ab}\) includes the out-of plane response of the current sheet \(\chi^s_{z}\) through \(\varphi_z\), and can therefore be compared to the thin film model.

\subsection{2D material as a thin film}

We present in this section the propagation in the effective thin film system of thickness \(d_f\) described by the diagonal tensor \(\tens{\varepsilon}^f\) of components  \(\varepsilon^f_x\), \(\varepsilon^f_y\), and \(\varepsilon^f_z\).  The total transfer matrix of the thin film (\(\mathcal{T}_{ab}\)) involves the transfer matrix at the two interfaces (\(\mathcal{I}_{af}\) and \(\mathcal{I}_{fb}\)) and the propagation matrix \(\mathcal{P}_{f}\) in the homogeneous film \(f\) over a distance \(d_f\). Then
\begin{equation}
\mathcal{T}_{ab} = \mathcal{I}_{af} \mathcal{P}_{f}\mathcal{I}_{fb}
\end{equation}
with
\begin{eqnarray}
\mathcal{P}_m     &=&  \left(\begin{array}{cc}\mathrm{e}^{-\mathrm{i}\Phi_m}&0\\0&\mathrm{e}^{\mathrm{i}\Phi_m}\end{array}\right),\\
\mathcal{I}_{mn}  &=&  \frac{1}{t_{mn}}\left(\begin{array}{cc}1&r_{mn}\\r_{mn}&1\end{array}\right),
\end{eqnarray}
where \(\Phi_m=k^m_z d_m\), \(d_m\) is the thickness of layer \(m\), and \(k^m_z\), \(r_{mn}\) and \(t_{mn}\) are defined in table~\ref{Tab:TETM}. The anisotropy of the media is described through the diagonal components of the dielectric tensors.

\subsection{Analytical comparison}

The two models are considered equivalent if their transfer matrices are identical. However, we cannot directly compare \(\mathcal{S}_{ab}\) and \(\mathcal{T}_{ab}\) since the propagation in the slab of thickness \(d_f=d_a + d_b\) is not considered in  \(\mathcal{S}_{ab}\). The correct equality is then
\begin{equation}
 \mathcal{T}_{ab} = \mathcal{P}_{a} \mathcal{S}_{ab} \mathcal{P}_{b}
 .\label{Eq:2Dequals3D}
\end{equation}

In the limit of small phase shift, we can develop (\ref{Eq:2Dequals3D}) to the first order in \(k_0 d_f\) for the bulk parameters (\(\Phi_a,\Phi_b,\Phi_f\ll1\)), and to the first order in \(k_0\chi^s\) for the single sheet parameters (\(\varphi_x,\varphi_y,\varphi_z\ll1\)). A lengthy but straightforward calculation provides the effective dielectric function of the thin film as
\begin{eqnarray}
 \varepsilon^f_x &=& \chi^s_{x}/d_f +\eta_a\varepsilon^a_x + \eta_b\varepsilon^b_x,\label{Eq:chixx}\\
 \varepsilon^f_y &=& \chi^s_{y}/d_f +\eta_a\varepsilon^a_y + \eta_b\varepsilon^b_y ,\label{Eq:chiyy}\\
 \frac{1}{\varepsilon^f_z} &=& \frac{\eta_a}{\varepsilon^a_z} + \frac{\eta_b}{\varepsilon^b_z} -\frac{\chi^s_{z}}{\epsilon_{ab}\,d_f},\label{Eq:chizz}
\end{eqnarray}
where \(\eta_a=d_a/d_f\), and \(\eta_b=d_b/d_f\) and then \(\eta_a+\eta_b=1\) (Fig.~\ref{Fig:interface}). As expected, the effective dielectric tensor components do not depend on the angle of incidence angle. Nevertheless, those quantities depend on the 2D material through \(\chi^s\), and on the geometry of the thin film defined by \(d_a\) and \(d_b\). More surprisingly, the components of the dielectric tensor of the surrounding materials \(\tens{\varepsilon}^a\) and \(\tens{\varepsilon}^b\) also appear. In the frequent case where the incident medium is air, and the thin film of thickness \(d_f\) is lying on top of the substrate \(b\), we have \(d_a=d_f\), \(d_b=0\) and \(\varepsilon^a_{i}=1\), so that
 \begin{eqnarray}
  \varepsilon^f_x &=& \chi^s_{x}/d_f + 1,\label{Eq:chixx_vaccuum}\\
  \varepsilon^f_y &=& \chi^s_{y}/d_f + 1,\label{Eq:chiyy_vaccuum}\\
  \frac{1}{\varepsilon^f_z} &=& 1  + \frac{1+\varepsilon^b_z}{2\varepsilon^b_z{d_f}}\chi^s_{z}. \label{Eq:chizz_vaccuum}
 \end{eqnarray}
Equations (\ref{Eq:chixx_vaccuum}) and~(\ref{Eq:chiyy_vaccuum}) are commonly used for 2D materials and perfectly valid under the assumptions reported above. The relation for the out-of-plane components, (\ref{Eq:chizz}) and~(\ref{Eq:chizz_vaccuum}), are far from being intuitive but are
important to understand the link between the isotropic thin film and the anisotropic single sheet models. Indeed, they explain some discrepancies between the two approaches reported in the literature, as we will discuss later. In the absence of out-of-plane susceptibility (\(\chi^s_{z}=0\)), (\ref{Eq:chizz_vaccuum}) gives \(\varepsilon^f_z=1\) and the effective thin film is anisotropic.

\section{Discussion}
In the following section we compare the results of the two approaches on quantities that are easily obtained experimentally: transmittance, ellipsometry and optical contrast.

\subsection{Transmittance}
In TM configuration, from (\ref{Eq:t}) and table~\ref{Tab:TETM}, the change in transmittance induced by the 2D material in the small phase shift hypothesis (first order in  \(k_0 d_f\sim{}k_0\chi^s\)) and for real \(\tens{\varepsilon}_a\) and \(\tens{\varepsilon}_b\) is
\begin{eqnarray}
\lefteqn{\left\vert\frac{t}{t_{ab}}\right\vert^2-1
}\nonumber\\
&=&
-t_{ab}
\frac{k^b_z}{\varepsilon^b_z}\mathrm{Im}{\chi^s_{x}}
\left[1+\frac{k_x^2}{k^a_zk^b_z}\frac{\varepsilon^a_x\varepsilon^b_z}{\epsilon_{ab}}
\left(\frac{\mathrm{Im}{\chi^s_{z}}}{\mathrm{Im}{\chi^s_{x}}}\right)\right]
\label{Eq:red_trans}\\
&=&
-t_{ab}\frac{k^b_z}{\varepsilon^b_z}\mathrm{Im}{\varepsilon^f_x}d_f
\left[
1
+\frac{k_x^2}{k^a_zk^b_z}\frac{(\varepsilon^a_x)^2\varepsilon^b_z}{\vert\varepsilon^f_z\vert^2}
\left(\frac{\mathrm{Im}{\varepsilon^f_z}}{\mathrm{Im}{\varepsilon^f_x}}\right)
\right].\nonumber\\
\label{Eq:red_transf}
\end{eqnarray}
The change in transmittance (\ref{Eq:red_trans}) is then only related to \(\mathrm{Im}\,\chi^s\) and, via (\ref{Eq:sigmas}), to \(\mathrm{Re}\,\sigma^s\). Simple transmittance measurements can therefore not provide information on \(\mathrm{Re}\,\chi^s\) or \(\mathrm{Im}\,\sigma^s\). In contrast, both the real and the imaginary parts of \(\varepsilon^f_z\) are present in (\ref{Eq:red_transf}) through \(\vert\varepsilon^f_z\vert^2\).

To understand the impact of using the thin film model instead of the single sheet approach, and to test the validity range of (\ref{Eq:chixx})--(\ref{Eq:chizz}) with respect to \(k_0\chi^s\), we have performed extensive numerical simulations. All the numerical results presented here are for a TM wave incident on an air/2D/glass (\(n_b=1.5\)) system with an angle \(\theta=75\)\textdegree, and a thickness \(d_f=0.34\)\,nm, except otherwise specified.

\begin{figure*}\centering
\begin{tabular}{cc}
\includegraphics[width=0.9\linewidth]{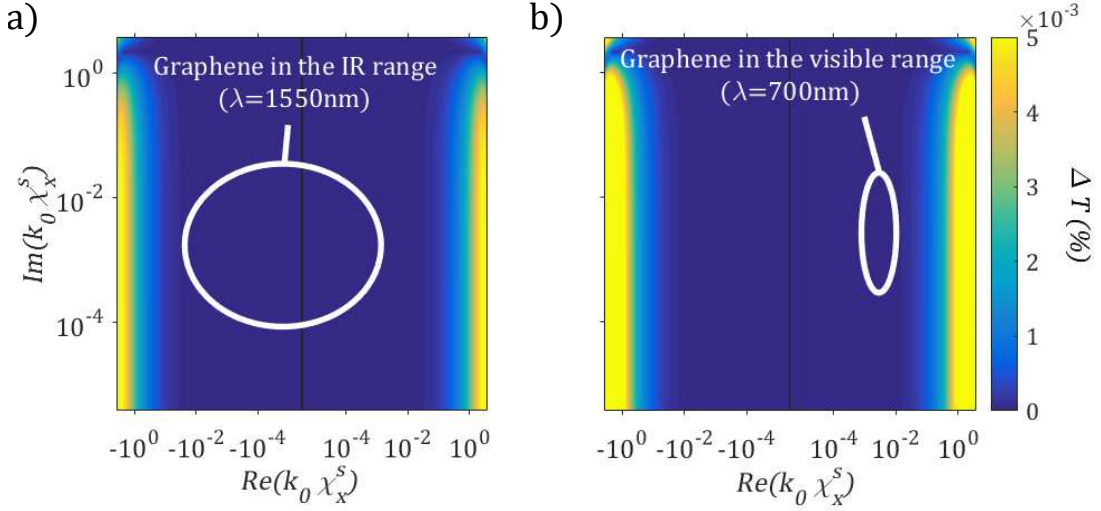}&%
\end{tabular}
\caption{Relative difference  \(\Delta T\) between the transmittance computed in the single sheet and in the anisotropic thin film models, with respect to the real and imaginary part of the single sheet susceptibility (\(k_0\chi_x^s\)).
The system considered is air/graphene/glass. The refractive index of glass is taken as \(1.5\).
(a) Infrared EM radiation (\(\lambda=1550\)\,nm); (b) Visible light (\(\lambda=700\)\,nm). Circled areas indicate the range of values \(k_0\chi_x^s\) for graphene within the Kubo formula for a range of Fermi level from \(0.05\)\,eV to \(1\)\,eV and a range of relaxation time from \(10\)\,fs to \(200\)\,fs.}
\label{Fig:2DvsFPa}
\end{figure*}

Fig.~\ref{Fig:2DvsFPa} displays the difference of transmittance (\(\Delta T\)) for two incident wavelengths (in the IR, \(\lambda=1550\)\,nm, and in the visible, \(\lambda=700\)\,nm) obtained with the single sheet model with no out-of-plane susceptibility (\(\chi^s_{z}=0\)) and with the anisotropic thin film with \(\tens{\varepsilon}^f\) from (\ref{Eq:chixx})-(\ref{Eq:chizz}). The transmittance computed in the \emph{anisotropic} thin film model and in the single sheet model are obviously in very good agreement. Therefore, in the following, we will consider that the single sheet model and the anisotropic thin film model give equivalent results. This rationalizes also the fact that the small phase shift condition is satisfied for a large range of 2D susceptibilities.

When the small phase shift condition is relaxed  (\(\left\vert\mathrm{Re}\left[k_0\chi^s\right]\right\vert\gtrapprox1\)), a small discrepancy can be observed, corresponding to the yellow bands on the sides of Fig.~\ref{Fig:2DvsFPa}(a) and (b). A comparison between these two figures shows that larger discrepancies are  observed at \(\lambda=700\)\,nm, than at \(\lambda=1550\)\,nm. For a better interpretation of the data, we identify on the figure possible values for graphene conductivity based on the Kubo formula~\cite{doi:10.1088/1742-6596/129/1/012004}. In this case, we observe a particularly small \(\Delta{T}\) with maximum of \(5\cdot10^{-3}\%\).

\begin{figure*}\centering
\begin{tabular}{cc}
\includegraphics[width=0.9\linewidth]{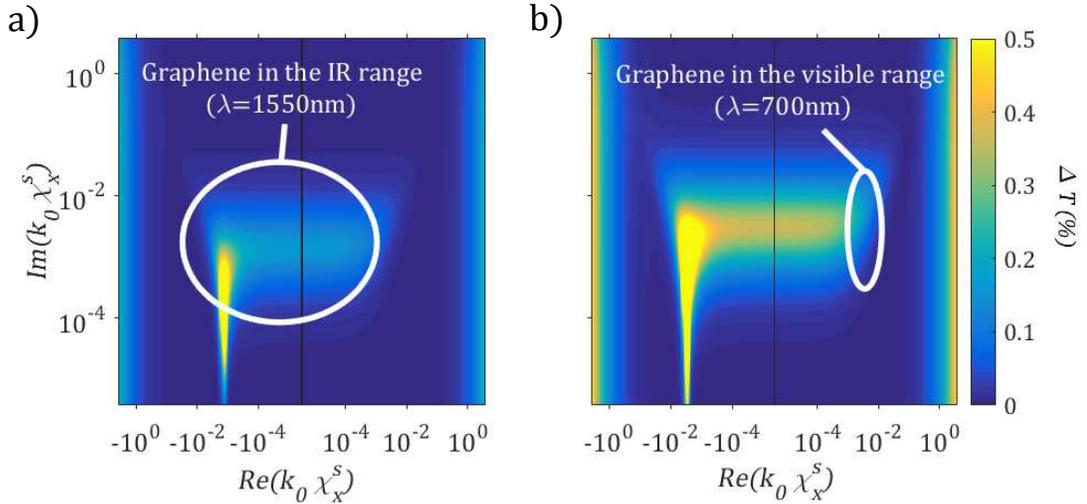}&%
\end{tabular}
\caption{Same as Fig.~\ref{Fig:2DvsFPa} for the isotropic thin film model.}
\label{Fig:2DvsFPi}
\end{figure*}

Interestingly, an isotropic thin film model shows more important discrepancies when compared with an anisotropic thin film, as illustrated on Fig.~\ref{Fig:2DvsFPi} even if, for a large range of values, both isotropic and anisotropic thin film models provide very similar results.
Note that the scale of \(\Delta T\) is different on Fig. \ref{Fig:2DvsFPa} and  Fig.~\ref{Fig:2DvsFPi}. For Fig.~\ref{Fig:2DvsFPi}, \(\varepsilon_{x}\) and \(\varepsilon_{y}\) are equal and obtained from (\ref{Eq:chixx}),(\ref{Eq:chiyy}) and \(\varepsilon_{x}= \varepsilon_{y}= \varepsilon_{z}\).

Notably, a high value of \(\Delta T\) is observed on a vertical line corresponding to \(\mathrm{Re}\left[\chi^s\right]=d_f\), for which the real part of the permittivity
\(
\varepsilon=1-\chi^s/d_f
\)
vanishes. This shows, similarly to what was reported in~\cite{Valuev2016}, that an artificial plasmonic resonance is predicted by an \emph{isotropic} thin film model, due to the artificial metallic nature of the out-of-plane component of the permittivity tensor. This unphysical resonance could have dramatic effects on the prediction of the optical properties.

\begin{figure}
    \centering
    \includegraphics[width=0.90\linewidth]{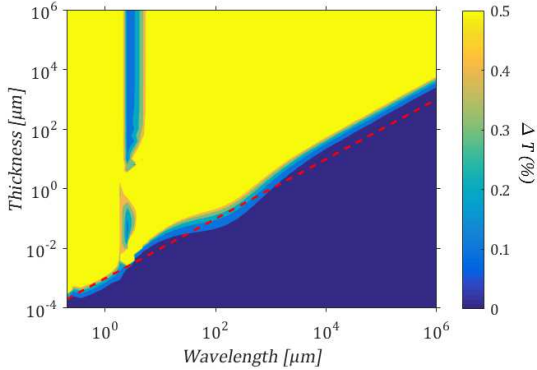}
    \caption{Difference of transmittance \(\Delta T\) between the isotropic and anisotropic thin film of graphene
    for a system air/graphene/glass. The refractive index of glass is taken as \(1.5\). Graphene is modeled using the Kubo formula with \(E_F=0.4\)\,eV, \(\tau=100\)\,fs. The red dotted line represent a thickness equal to \(1/1000\) of the wavelength.}
    \label{Fig:3Diso3Dani}
\end{figure}

To investigate further the influence of the anisotropy, we present in Fig.~\ref{Fig:3Diso3Dani} the difference between the transmittance obtained with the isotropic and anisotropic thin film models in TM configuration for graphene, as a function of the incident wavelength and of the thin film thickness. The two models give very similar results for a ratio \(\lambda / d_f > 1000\) (dashed line) (\textit{i.e.} very small \(k_0d_f\)). This validates the fact that the anisotropy of graphene has been often disregarded without consequences on the validity of the conclusions.

This surprisingly good predictions within the isotropic thin film model for intrinsically anisotropic 2D material is explained as follows. The isotropic thin film model does not correspond to the assumption \(\chi^s_{z}=\chi^s_{x}\), but to \(\varepsilon^f_z=\varepsilon^f_x\). By means of (\ref{Eq:chizz}), this is equivalent to set
\begin{equation}
\mathrm{Im}\chi^s_{z}
=\frac{\varepsilon^a_x\epsilon_{ab}}{\vert{\varepsilon^f_z}\vert^2}
d_f\,\mathrm{Im}\varepsilon^f_{z}
=\frac{\varepsilon^a_x\epsilon_{ab}}{\vert{}1+\chi^s_{x}/d_f\vert^2}
\mathrm{Im}\chi^s_{x}
\label{Eq:chizziso3D}
,
\end{equation}
which shows that the isotropic thin film model tends to the \emph{anisotropic} single sheet one with \(\chi^s_{z}=0\) when \(\vert{\chi^s_{x}}\vert^2\gg{}d_f^2\), as in this case (\ref{Eq:chizziso3D}) provides \(\chi^s_{z}\approx0\).

\begin{figure}\centering
\includegraphics[totalheight=6cm]{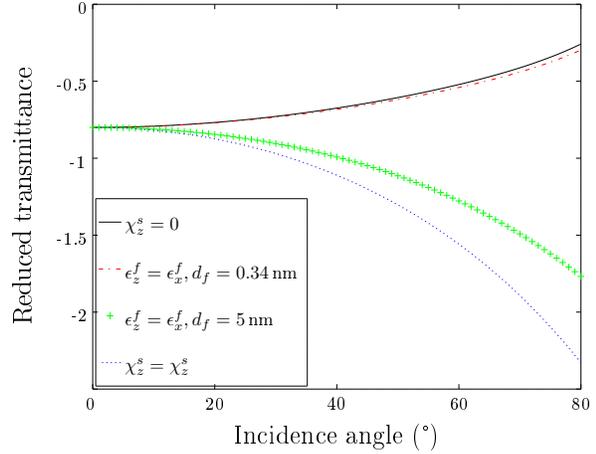}
\caption{%
Reduced transmittance [brackets in (\ref{Eq:red_trans}) and (\ref{Eq:red_transf})]
   for and air/graphene/glass structure:
   \(n_a=1\), \(n_b=1.5\), \(\chi^s_{x}=(1.50+2.29\mathrm{i})\,\mathrm{nm}\) at a wavelength of  \(634\)\,nm~\cite{cheon_how_2014}.
   Curves are for:
   anisotropic single sheet with no out-of-plane response (\(\chi^s_{z}=0\));
   isotropic thin film model (\(\varepsilon^f_z=\varepsilon^f_x\))
                             with \(d_f=0.34\)\,nm
                             and \(d_f=5\)\,nm;
   isotropic current sheet model (\(\chi^s_{z}=\chi^s_{x}\)).}
\label{Fig:transmit2D3D}
\end{figure}

This justifies that the isotropic thin film model can be used with good results to model graphene with \(\mathrm{Im}[\chi^s_{z}]=0\), and \(d_f=0.34\)\,nm. In particular, it confirms that in the limit \(d_f=0\), both models agree, as reported in~\cite{10.1088/978-1-6817-4309-7}.
More importantly, this also resolves the apparent contradiction between the conclusions of ~\cite{10.1088/978-1-6817-4309-7}, and those of~\cite{PhysRevA.93.013832,Matthes_16,Valuev2016} on the equivalence (or not) of both models for \(d_f\rightarrow0\).
Indeed, (\ref{Eq:chizziso3D}) shows that, if we model graphene by means of a thicker layer so that \(\chi^s_{x}\ll{d_f}\ll1/k_0\), the isotropic thin film model corresponds to the \emph{isotropic} single sheet one (\(\chi^s_{x}=\chi^s_{z}\) in the particular case were \(n_a=n_b=1\)) and no more to the \emph{anisotropic} single sheet with \(\chi^s_{z}=0\). We illustrate this analytical observation on Fig.~\ref{Fig:transmit2D3D}, where we plot the bracket in (\ref{Eq:red_trans}) and~(\ref{Eq:red_transf}) for an air/graphene/glass system at 634\,nm as function of the angle of incidence for different approaches:
the single sheet for \(\chi^s_{z}=0\) (full line) and with (\(\chi^s_{x}=\chi^s_{z}\)) (dashed line), the isotropic thin film for \(d_f=0.34\)\,nm (dot-dashed line) and \(d_f=5\)\,nm (crosses). The results for the anisotropic thin film cannot be distinguished from those of the anisotropic single sheet, and are therefore not plotted. The thickness \(d_f=5\)\,nm is commonly used in discrete numerical simulations to avoid prohibitive numerical cost~\cite{Valuev2016}. As expected, at normal incidence, all curves are superimposed and the anisotropy does not play any role. As the angle of incidence increases, the \(z\)-component of the response functions becomes more important and the exact value of the thickness of the thin film influences the computed optical properties.

Although transmittance change at different angles would in principle allow to separate the in-plane and the out-of-plane responses of the imaginary part of the susceptibility (real part of the conductivity), these measurements are usually performed at normal incidence, for which the TE and TM cases coincide.

\subsection{Ellipsometry}
Ellipsometry records the ratio of the reflexion or transmission of a sample in TM and TE configurations, at different angles.  From equations (\ref{Eq:t}) and (\ref{Eq:r}), still in the small phase approximation (\textit{i.e.} to the first order in \(k_0\chi^s\)), we get
\begin{equation}
\frac{\rho^t}{\rho^t_0}=\frac{t^{TM}}{t^{TM}_{ab}}\frac{t^{TE}_{ab}}{t^{TE}} =
 \left[1+\mathrm{i}\left(\varphi^{TM}_x-\varphi^{TE}_y+\varphi^{TM}_z\right)\right],
\end{equation}
with \(\rho^t\) the ellipsometric ratio with the 2D material, and \(\rho^t_0\) the same ratio for the interface without 2D material.
Under the assumption that \(\chi^s_{x}=\chi^s_{y}\), using table~\ref{Tab:TETM}, we can write
\begin{equation}
  \frac{\rho^t}{\rho^t_0}-1
 =
 -\frac{\mathrm{i}k_x^2}{k^a_z+k^b_z}
 \left(\chi^s_{x}-\frac{\varepsilon^a_x\varepsilon^b_x}{\epsilon_{ab}}\chi^s_{z}\right).
\label{Eq:ellips1}
\end{equation}
The in-plane \(\chi^s_{x}\) and the out-of plane \(\chi^s_{z}\) susceptibilities can then not be separated by means of standard transmission ellipsometry as the coefficient in front of \(\chi^s_z\) is independent of \(k_x\) and \(k_z\), and therefore of the angle of incidence.

\begin{figure}[h!]\centering
\includegraphics[totalheight=6cm]{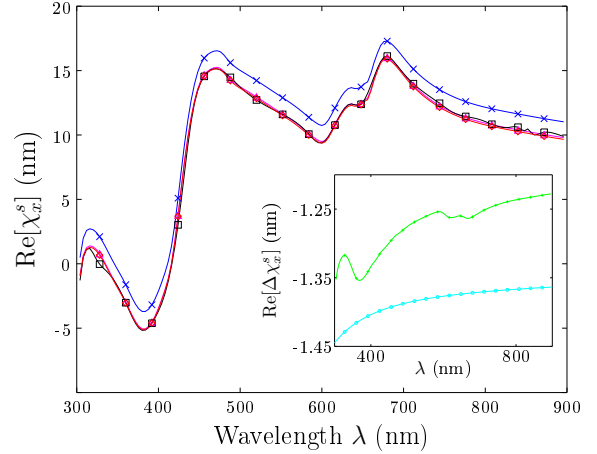}
\caption{%
Values of \(\mathrm{Re}[\chi^s_x]\) for \MoS2 single layer retrieved with the isotropic thin film model (\textcolor{blue}{x}) and the anisotropic current sheet model (\textcolor{black}{\(\square\)}). Data from~\cite{Jayaswal_18}. The (\textcolor{red}{\(\diamond\)}) curve is calculated from (\textcolor{blue}{\(\times\)}) using the shift calculated from (\ref{Eq:chia_from_chiI}) (\textcolor{green}{\(+\)} in the inset);
(\textcolor{magenta}{\(\Delta\)}) is (\textcolor{blue}{x}) shifted by \(-\varepsilon^b_x d\) (\textcolor{cyan}{\(o\)} in the inset). The dielectric function of N-BK7 glass (substrate used in~\cite{Jayaswal_18}) is taken from Sellmeier's equation provided by Schott~\cite{rii-BK7}.}
\label{Fig:MoS2-iso-vs-anis}
\end{figure}
In order to analyze further these results, it is convenient to define the parenthesis in (\ref{Eq:ellips1}) as
\begin{equation}
 \chi^{ell}=\chi^s_{x}-\frac{\varepsilon^a_x\varepsilon^b_x}{\epsilon_{ab}}\chi^s_{z},\label{Eq:chiell}
\end{equation}
that contains all the dependence in the susceptibility. Note that the same dependence is found for reflexion ellipsometry.

We can here provide a simple explanation to the difference reported in~\cite{Jayaswal_18} between the susceptibilities extracted from ellipsometric data using different models for \MoS2 on glass substrate. We named \([\chi^s_{x}]_{i}\) the in-plane susceptibility deduced from the isotropic thin film model with \(d_f=0.615\) nm (Fig. \ref{Fig:MoS2-iso-vs-anis}, blue curve with \(\times\)) and \([\chi^s_{x}]_{a}\) the one found with the anisotropic single sheet model (Fig. \ref{Fig:MoS2-iso-vs-anis}, red curve with \(\square\) ).

For the isotropic thin film, using~(\ref{Eq:chizz_vaccuum}),
\begin{equation}
 \chi^{ell}
           =[\chi^s_{x}]_{i}
            \left(1-\frac{\varepsilon^b_x}{1+[\chi^s_{x}]_{i}/d_f}\right),
\label{Eq:chiell_i}
\end{equation}
while for the anisotropic sheet, we simply get
\begin{equation}
 \chi^{ell} =[\chi^s_{x}]_{a}.
 \label{Eq:chiell_a}
\end{equation}
Comparing the last two equations, we obtain
\begin{equation}
[\chi^s_{x}]_{a}=[\chi^s_{x}]_{i} -\varepsilon^b_x d_f
                  + \frac{\varepsilon^b_x d_f}{1+[\chi^s_{x}]_{i}/d_f}
.\label{Eq:chia_from_chiI}
\end{equation}

The inset of Fig. \ref{Fig:MoS2-iso-vs-anis} displays the difference between \([\chi^s_{x}]_{i}\) and \([\chi^s_{x}]_{a}\) and compares it with \(-\varepsilon^b_x d_f\). We see that the last term in (\ref{Eq:chia_from_chiI}) is negligible.  We do not reproduce the imaginary part of the susceptibility (real part of the conductivity), as it is not affected by the real shift \(-\varepsilon^b_x d_f\), as is visible in~\cite{Jayaswal_18}.

This good agreement between experimental data and the analytical predictions again confirms the validity of the small phase shift approximation (\(k_0 \chi^s\ll1\)).

We conclude  that  standard ellipsometry provides no information on the \(x-z\) anisotropy of the 2D sample.
It is however important to note that the single sheet approach imposes implicitly \(\chi^s_{z}=0\), while the isotropic thin film approach assumes \(\chi^s_{z}=\epsilon_{ab}\chi^s_{x}/\varepsilon^f_x\), which explains differences reported in the literature, for the retrieval of \(\chi^s_{x}\) from ellipsometric data. Equation (\ref{Eq:chiell}) allows to introduce in the fitting procedure a value for \(\chi^s_{z}\) based on theoretical assumptions or obtained experimentally, for example, by means of contrast ratio measurements.

\subsection{Optical contrast}

The optical contrast of 2D materials on a thick substrate is often very small and hardly measurable. However, reflexion microscopy and optical contrast are commonly used to determine the presence of 2D materials (or the number of layers) if a thin dielectric film is lying on top of the substrate, most often \SiO2 on \Si~\cite{Bayle_18,doi:10.1063/1.2768625,doi:10.1063/1.2768624}. This additional layer creates interferences that depend on the 2D susceptibility and allow to tune the total reflectance of the system. Measurements are usually performed at normal incidence, so that only the in-plane susceptibility is probed.
To take into account the additional layer, we should simply multiply \(\mathcal{T}_{ab}\) in (\ref{Eq:2Dequals3D}) by a propagation matrix accounting for the propagation in the additional layer, and an interface matrix between this layer and the substrate.
As the matrix \(\mathcal{T}_{ab}\) is the same in the current sheet and thin film approaches, the final result do not depend on the chosen model, especially at normal incidence for which the isotropic and anisotropic thin film models are equivalent.

The optical contrast then depends on the real and imaginary parts of the in-plane susceptibility of the 2D material. However, this measurement is strongly dependent on the parameters of the top layer, including their thickness and permittivity. This explains probably the differences in fitting experimental data that were reported in~\cite{PhysRevA.93.013832}.

\section{Conclusions}

We have explored analytically and numerically the link between the description of a 2D material with a current sheet or thin film model. We have focused our analysis on the description of the intrinsic anisotropy of the layers, \textit{i.e.} the effect of the out-of-plane component of the single sheet susceptibility or the out-of-plane component of the dielectric tensor of the thin film.  The analytical equivalence in the small phase shift condition shows that most discrepancies between these two approaches do not come from the finite thickness of the thin film, but from an incorrect description of the anisotropy, mainly for TM polarization and oblique incidence. In particular, we have shown that considering an isotropic dielectric function of a thin film is not equivalent to assume an isotropic single sheet susceptibility. We have  also commented the fact that a single sheet with vanishing out-of-plane response (as graphene) corresponds to an isotropic or an anisotropic effective thin film depending of the effective thickness of the film.

The application of the transfer matrix approach to classical measurement schemes provides evidences that a combination between different techniques is needed to fully characterize a 2D material, as
\begin{itemize}
 \item transmittance measurements on dielectric substrate provide \(\mathrm{Im}[\chi^s_{x}]\) at normal incidence and could provide \(\mathrm{Im}[\chi^s_{z}]\) at other incidence angle;
 \item standard ellipsometry, in transmission or reflection, cannot separate the in-plane and out-of-plane contributions. However, combined with transmittance changes it could provide \(\mathrm{Im}[\chi^s_{z}]\). Another way to retrieve \(\chi^s_{x}\) and \(\chi^s_{z}\) separately would be to perform ellipsometry experiments on different substrates;
 \item optical contrast on a multilayer substrate combined with the previous methods can also provide information about \(\mathrm{Re}[\chi^s_{x}]\), or even \(\mathrm{Re}[\chi^s_{z}]\) at oblique incidence.
\end{itemize}

We hope that the present single sheet transfer matrix approach, and the analytical connection with the thin film model will help to efficiently perform the analysis of stratified media involving 2D materials, either for the characterization of 2D materials, or for their use in various applications.

\section*{Acknowledgments}

Part of this research was performed while E. D. was
funded by the Fund for Research Training in Industry
and Agriculture (FRIA, Belgium) and the ``Fonds Van Buuren/Jaumotte-Demoulin'', and M. L. was a recipient of
a Fellowship of the Belgian American Educational Foundation.
This research used resources of the ``Plateforme Technologique de Calcul Intensif (PTCI)''
(http://www.ptci.unamur.be) located at the University of Namur, Belgium, which is supported
by the F.R.S.-FNRS under the convention No. 2.5020.11. The PTCI is member of the ``Consortium des Équipements de Calcul Intensif (CÉCI)'' (http://www.ceci-hpc.be).


\begin{thebibliography}{10}
\newcommand{\enquote}[1]{``#1''}

\bibitem{doi:10.1002/adma.201606128}
S.~Yu, X.~Wu \emph{et~al.}, Advanced Materials \textbf{29}, 1606128.

\bibitem{doi:10.1038/nphoton.2010.186}
F.~Bonaccorso, Z.~Sun \emph{et~al.}, Nature Photonics \textbf{4}, 611 (2010).

\bibitem{doi:10.1039/c4nr01600a}
A.~C. Ferrari, F.~Bonaccorso \emph{et~al.}, Nanoscale \textbf{7}, 4598 (2015).

\bibitem{doi:10.1038/nphoton.2015.282}
K.~F. Mak and J.~Shan, Nature Photonics \textbf{10}, 216 (2016).

\bibitem{Bayle_18}
M.~Bayle, N.~Reckinger \emph{et~al.}, Journal of Raman Spectroscopy
  \textbf{49}, 36.

\bibitem{doi:10.1063/1.2768625}
D.~S.~L. Abergel, A.~Russell, and V.~I. Fal’ko, Applied Physics Letters
  \textbf{91}, 063125 (2007).

\bibitem{Eichfeld_2014}
S.~M. Eichfeld, C.~M. Eichfeld \emph{et~al.}, APL Materials \textbf{2}, 092508
  (2014).

\bibitem{Li_2013}
H.~Li, J.~Wu \emph{et~al.}, ACS Nano \textbf{7}, 10344 (2013). PMID: 24131442.

\bibitem{ottaviano_mechanical_2017}
L.~Ottaviano, S.~Palleschi \emph{et~al.}, 2D Materials \textbf{4}, 045013
  (2017).

\bibitem{PhysRevB.78.085432}
T.~Stauber, N.~M.~R. Peres, and A.~K. Geim, Phys. Rev. B \textbf{78}, 085432
  (2008).

\bibitem{Lobet_16}
M.~Lobet, B.~Majerus \emph{et~al.}, Phys. Rev. B \textbf{93}, 235424 (2016).

\bibitem{doi:10.1088/1742-6596/129/1/012004}
L.~A. Falkovsky, Journal of Physics: Conference Series \textbf{129}, 012004
  (2008).

\bibitem{Majerus_18}
B.~Majérus, M.~Cormann \emph{et~al.}, 2D Materials \textbf{5}, 025007 (2018).

\bibitem{Nelson_2010}
F.~J. Nelson, V.~K. Kamineni \emph{et~al.}, Applied Physics Letters
  \textbf{97}, 253110 (2010).

\bibitem{cheon_how_2014}
S.~Cheon, K.~D. Kihm \emph{et~al.}, Scientific Reports \textbf{4}, 6364 (2014).

\bibitem{Lu_97}
J.~P. Lu, Phys. Rev. Lett. \textbf{79}, 1297 (1997).

\bibitem{Wagner_13}
P.~Wagner, V.~V. Ivanovskaya \emph{et~al.}, Journal of Physics: Condensed
  Matter \textbf{25}, 155302 (2013).

\bibitem{PhysRevA.93.013832}
M.~Merano, Phys. Rev. A \textbf{93}, 013832 (2016).

\bibitem{Matthes_16}
L.~Matthes, O.~Pulci, and F.~Bechstedt, Phys. Rev. B \textbf{94}, 205408
  (2016).

\bibitem{Jayaswal_18}
G.~Jayaswal, Z.~Dai \emph{et~al.}, Opt. Lett. \textbf{43}, 703 (2018).

\bibitem{Valuev2016}
I.~Valuev, S.~Belousov \emph{et~al.}, Applied Physics A \textbf{123}, 60
  (2016).

\bibitem{Li_Heinz_18}
Y.~Li and T.~F. Heinz, 2D Materials \textbf{5}, 025021 (2018).

\bibitem{Funke_16}
S.~Funke, B.~Miller \emph{et~al.}, Journal of Physics: Condensed Matter
  \textbf{28}, 385301 (2016).

\bibitem{Kravets_10}
V.~G. Kravets, A.~N. Grigorenko \emph{et~al.}, Phys. Rev. B \textbf{81}, 155413
  (2010).

\bibitem{Marinopoulos_04}
A.~G. Marinopoulos, L.~Reining \emph{et~al.}, Phys. Rev. B \textbf{69}, 245419
  (2004).

\bibitem{PhysRevB.96.235422}
E.~Dremetsika and P.~Kockaert, Phys. Rev. B \textbf{96}, 235422 (2017).

\bibitem{Novoselov_16}
K.~S. Novoselov, A.~Mishchenko \emph{et~al.}, Science \textbf{353} (2016).

\bibitem{Rasmussen_15}
F.~A. Rasmussen and K.~S. Thygesen, The Journal of Physical Chemistry C
  \textbf{119}, 13169 (2015).

\bibitem{azzam1987ellipsometry}
R.~M.~A. Azzam, \emph{Ellipsometry and polarized light} (North-Holland Sole
  distributors for the USA and Canada, Elsevier Science Pub. Co, Amsterdam New
  York, 1987).

\bibitem{sipe_new_1987}
J.~E. Sipe, J. Opt. Soc. Am. B, JOSAB \textbf{4}, 481 (1987).

\bibitem{felderhof_electromagnetic_1987}
B.~U. Felderhof and G.~Marowsky, Appl. Phys. B \textbf{44}, 11 (1987).

\bibitem{10.1088/978-1-6817-4309-7}
R.~A. Depine, \emph{Graphene Optics: Electromagnetic Solution of Canonical
  Problems}, 2053-2571 (Morgan \& Claypool Publishers, 2016).

\bibitem{rii-BK7}
M.~N. Polyanskiy, \enquote{Refractive index database,}
  \url{https://refractiveindex.info}. Accessed on 2018-07-05.

\bibitem{doi:10.1063/1.2768624}
P.~Blake, E.~W. Hill \emph{et~al.}, Applied Physics Letters \textbf{91}, 063124
  (2007).

\end{thebibliography}
\end{document}